\documentclass[11pt]{article}
\usepackage{graphicx}
\usepackage{amsmath}
\usepackage{amssymb}
\usepackage{amsxtra}
\usepackage{float}

\usepackage{subcaption, csquotes}

\usepackage[
	style=phys,
	biblabel=brackets,
	eprint=true,
	bibencoding=auto,
	backend=biber,
	sorting=none,
	maxbibnames=3,
	language=auto,
	autolang=other,
	doi=false,
	url=false
	]{biblatex}

\AtEveryBibitem{
	\clearfield{day}
	\clearfield{month}
	\clearfield{endday}
	\clearfield{endmonth}
}

\usepackage[colorlinks]{hyperref}

\usepackage{authblk}

\title{Cosmological Formation of (2+1)-Dimensional Soliton Structures in Models Possessing Potentials with Local Peaks}
\author{A.~A.~Kirillov\thanks{\href{AAKirillov@mephi.ru}{AAKirillov@mephi.ru}}}
\author{B.~S.~Murygin\thanks{\href{MuryginBS@gmail.com}{MuryginBS@gmail.com}}}
\author{V.~V.~Nikulin\thanks{\href{VVNikulin@mephi.ru}{VVNikulin@mephi.ru}}}

\affil{
    National Research Nuclear University MEPhI
    \par
    (Moscow Engineering Physics Institute), 
    \par
    115409 Kashirskoe shosse 31, Moscow, Russia
    }

\date{}

\begin{document}
\maketitle

\begin{abstract}
    Production of domain walls and string-like solitons in the model with two real scalar fields and potential with at least one saddle point and a local maximum is considered. The model is regarded as 2-dimensional spatial slices of 3-dimensional entire structures. It is shown that, in the early Universe, both types of solitons may appear. In addition, the qualitative estimate of the domain walls and strings formation probabilities is presented. It is found that the probability of the formation of string-like solitons is suppressed compared to that of domain walls.
\end{abstract}

\noindent Keywords: solitons, strings, domain walls, early Universe

\noindent PACS: 03.50.-z; 11.27.+d; 98.80.Cq

\section{Introduction}
\label{sec:intro}

Nowadays, there are a lot of inflation theories describing evolution of scalar fields with potentials of both power and non-power forms; see, e.g.,~\cite{1993PhRvD..47..426A, 1994PhRvD..49..748L,  1998PhRvD..58f1301L, 2005JCAP...01..005K, 2013JCAP...07..038B, 2015JCAP...06..040P, 2015PhRvD..92b1303E}. Furthermore, string and supergravity theories predict the existence of a landscape of vacua~\cite{2003dmci.confE..26S}. Complicated relief forms of potentials could lead to a formation of solitons in the early Universe, such as domain walls or strings, even if the field in question is an inflaton one~\cite{1991PhRvD..44..340B, 2016JCAP...02..064G, 2017JCAP...04..050D, 2017JCAP...12..044D}. Moreover, it might affect the following evolution of the Universe if its abundance is significant enough~\cite{1985PhR...121..263V, vilenkin}. Collapse of domain walls and strings might lead to primordial black holes (PBHs) formation~\cite{1993PhRvD..47.3265G, 2000IJMPA..15.4433H, 2010JCAP...05..032H, 2018JCAP...11..008V, 2019PhRvD..99j4028H, 2020PhRvD.101b3513L} or even clusters of PBHs~\cite{2001JETP...92..921R, 2019EPJC...79..246B} having a lot of astrophysical evidences. Whether primordial black holes could significantly contribute to dark matter is a widely-discussed question~\cite{2020ARNPS..70..355C}.

In this paper, the~formation of solitons in fields models with potentials having at least one saddle point and a local maximum is studied. The~considered relief form often occurs in theories of modified multidimensional gravity~\cite{2007GrCo...13..253B}.
In addition, supergravity and string theories produce models with effective potentials which might be locally described by the potential form under consideration~\cite{2021Univ....7..115K}. Previously, solitons formation in the models with the same potential of two scalar fields in the (1 + 1)-space-time was considered~\cite{2017JPhCS.934a2046G, 2018JCAP...04..042G}, and the first results in the (2 + 1)-models were obtained~\cite{2020arXiv201107041K, 2020JPhCS1690a2144K}. 
 In this paper, the study of the (2 + 1)-model is continued. This configuration, leading to the formation of strings and walls during the inflation, is actually a 2-dimensional spatial slice of (3 + 1)-dimensional structures looking like ``pancakes'' or ``bubbles'' \cite{1998PhRvD..59b3505C}. As~it is shown below, inside these structures, strings and walls are connected.
Nevertheless, a string or a domain wall can be considered apart from the the entire structure. Such examination is made here, in order to study the stability of the solitons and to give qualitative estimates of the relative probabilities of the soliton production.

\section{Model}

Let us consider two real scalar fields $\varphi$ and $\chi$ with the Lagrangian
\begin{equation}
    \label{Lagrangian}
    \mathcal{L} = \frac{1}{2}g^{\mu\nu}
        \big(\partial_{\mu}\varphi\partial_{\nu}\varphi +
        \partial_{\mu}\chi\partial_{\nu}\chi\big) -
        \mathcal{V}(\varphi,\chi).
\end{equation}
Here, $g^{\mu\nu}=\text{diag}(1,-a^2(t),-a^2(t),-a^2(t))$ is the Friedman--Robertson--Walker metric tensor in the inflationary Universe, $\mathcal{V}$ is the model potential, $a(t)=e^{Ht}$ is the inflationary {scale factor, $t$ is the time from the beginning of inflation}, and $H \sim 10^{13}$~GeV is the Hubble parameter during the inflation. The indices, $\mu, \nu = 0, 1, 2, 3,$ are the indices for the time-space components.

The dimension of the physical space can be effectively reduced if there is (at least an approximate) symmetry. Solitons arising in 2-field potentials are extended in a certain direction~\cite{1995RPPh...58..477H}. Therefore, one can locally neglect the change in the fields in this direction (the $z$-axis is chosen as such direction), so $\varphi = \varphi(t,x,y)$. Thus, from Equation \eqref{Lagrangian}, one finds the two 2-dimensional field equations:
\begin{equation}
    \label{Motion}
    \begin{aligned}
        \varphi_{tt} + 3H\varphi_t-\varphi_{xx}-\varphi_{yy} &= -\frac{\partial\mathcal{V}}{\partial{\varphi}},
        \\
        \chi_{{t}{t}} + 3H\chi_t-\chi_{xx}-\chi_{yy} &= -\frac{\partial\mathcal{V}}{\partial\chi}.
     \end{aligned}
\end{equation}
The Hubble parameter gives the natural energy units for the model. Hereinafter, all variables are expressed in $H$ units.

In order to demonstrate the emergence and evolution of local string and wall configurations, one have to choose an appropriate mapping from the 
field $(\varphi,\chi)$-space to the physical $(x,y)$-space as an initial condition. During~the cosmological inflation, topologically similar mappings arise with a non-zero probability due to quantum fluctuations of the fields. Hence, for~simplicity, the~initial conditions for Equation \eqref{Motion} have been chosen as follows:
\begin{equation}
    \label{InitialConditions}
    \begin{aligned}
        \varphi(x,y,0) &=\mathcal{R}\cos{\Theta}+\varphi_1, & \varphi_t(x,y,0) &= 0; 
        \\
        \chi(x,y,0) &=\mathcal{R}\sin{\Theta}+\chi_1, & \chi_t(x,y,0) &= 0,
    \end{aligned}
\end{equation}
where
\begin{equation}
    \label{InitialConditionsLegend}
    \begin{aligned}
        \mathcal{R}(r) &= \mathcal{R}_0 \cosh^{-1}{\cfrac{r_0}{r}}, & &r\geq0;
        \\
        \Theta(\theta) &= \theta, & 0 \leq &\theta\leq 2\pi.
    \end{aligned}
\end{equation}

This mapping allows the disc-like initial fields distribution to cover the entire physical $xy$-plane. The~parameters $\varphi_1$ and  $\chi_1$ set the center of the disc-like distribution with the radius $\mathcal{R}_0$. The~parameters $r$ and $\theta$ represent the polar coordinates in the $xy$-plane. $r_0$ is introduced to make the ratio $r/r_0$ dimensionless.

The most natural boundary conditions are:
\begin{equation}
    \label{BoundaryConditions}
    \begin{aligned}
        \varphi_x(\pm\infty,y,t) & = 0, & \varphi_y(x,\pm\infty,t)& = 0;
        \\
        \chi_x(\pm\infty,y,t) & = 0, & \chi_y(x,\pm\infty,t)& = 0.
    \end{aligned}
\end{equation}
In the present calculations, the~potential is chosen in the form,
\begin{equation}
    \label{Potential1}
    \mathcal{V}(\varphi,\chi) = \cfrac{m^2}{2}(\varphi^2+\chi^2) +
        \Lambda^4 \exp\Big[-\lambda\big( (\varphi-\varphi_0)^2 + (\chi-\chi_0)^2 \big)\Big].
\end{equation}

 The potential approximates field models with a global minimum, a~saddle point and a 
local  maximum, mentioned in Section~\ref{sec:intro}.
Here, $m$ is the mass of each scalar. The~second term corresponds to the local 
maximum at the point $(\varphi_0, \chi_0)$. The~positive parameters $\Lambda$ and $\lambda$ describe the height and the width of the local maximum, respectively.

\section{Results}

\begin{figure}[H]
    \begin{subfigure}[h]{0.3\linewidth}
        \includegraphics[width=1\linewidth]{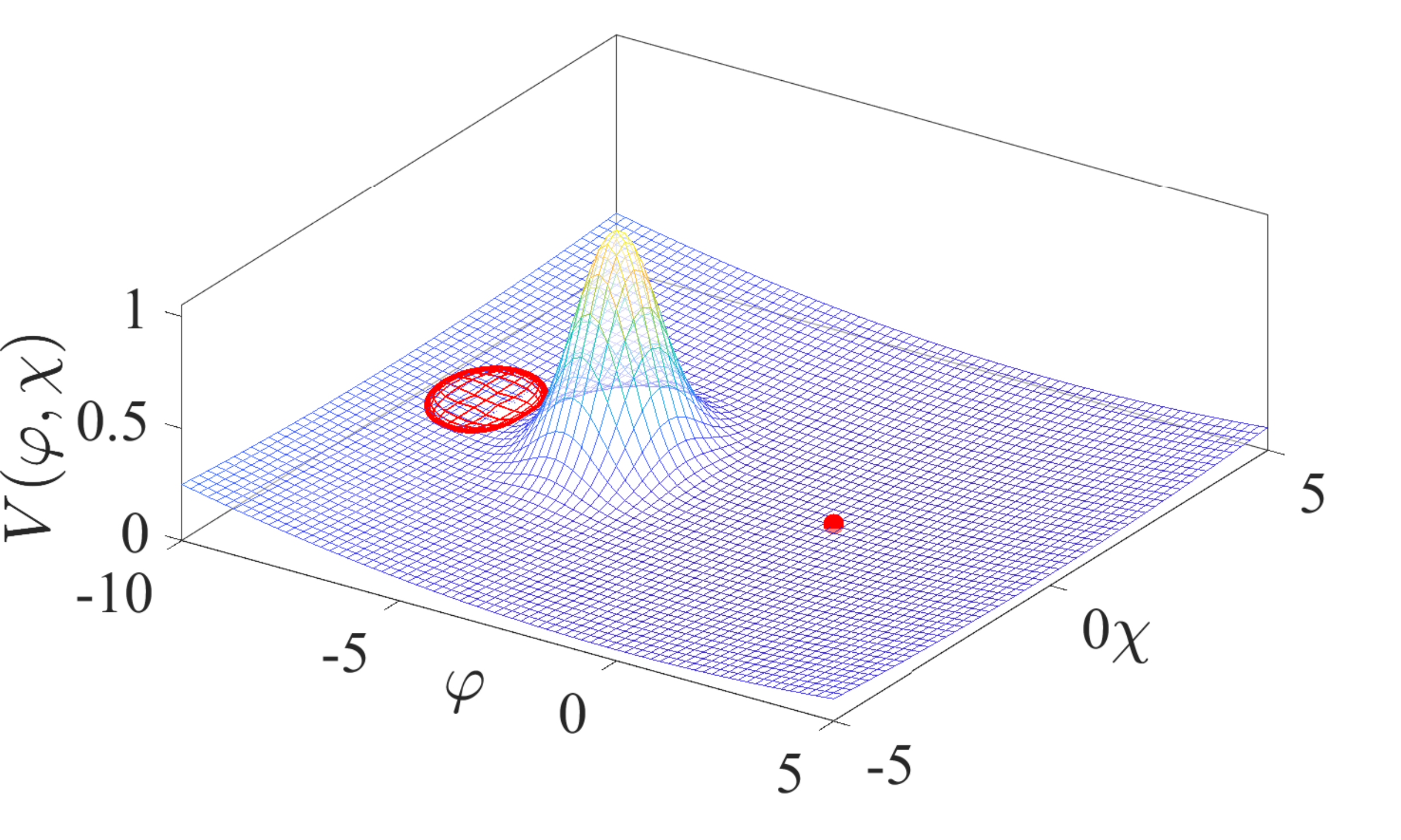} 
        \caption{}
        \label{fig:V_in_dw}
    \end{subfigure}
    \begin{subfigure}[h]{0.3\linewidth}
        \includegraphics[width=1\linewidth]{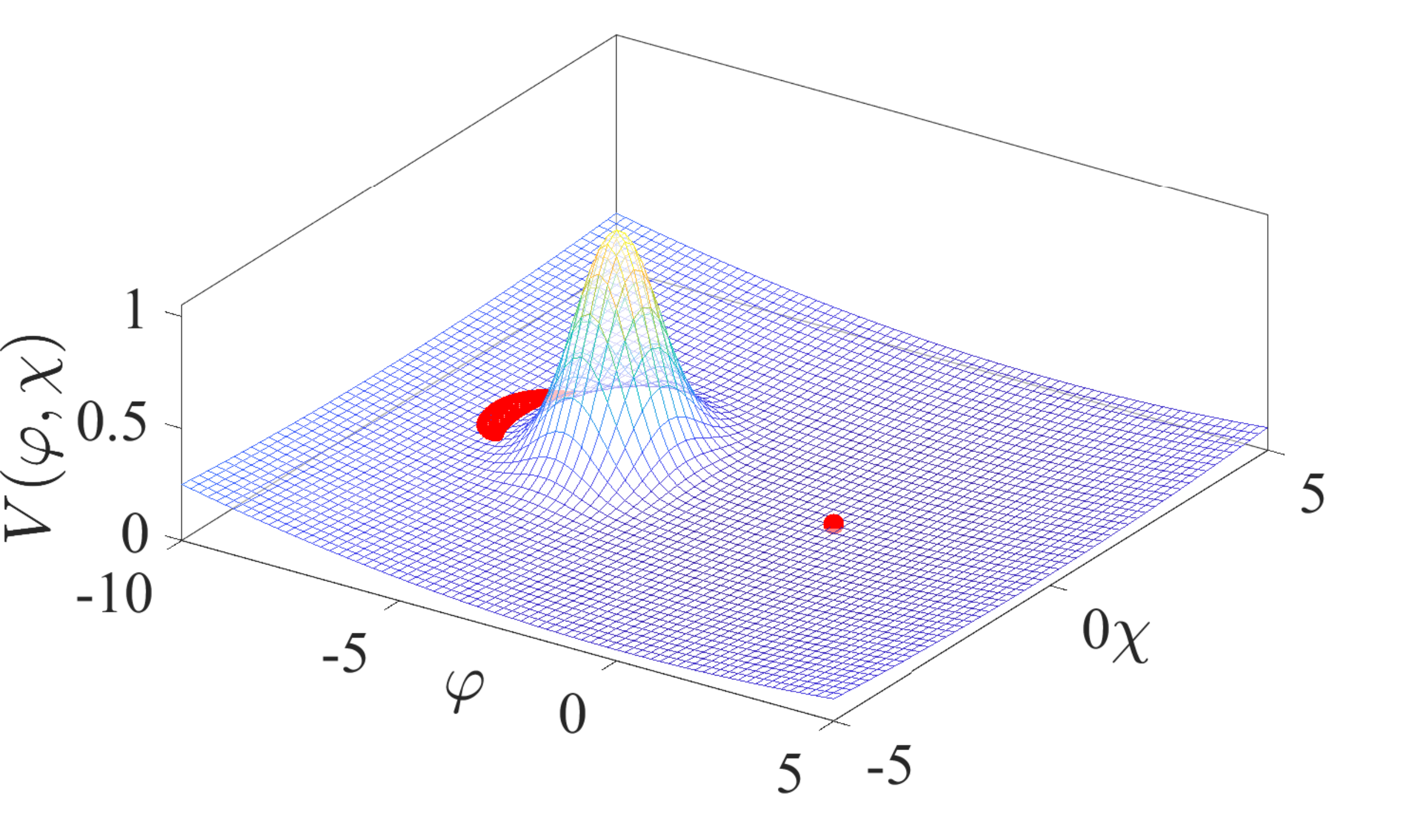}
        \caption{}
        \label{fig:V_inter_dw}
    \end{subfigure}
    \begin{subfigure}[h]{0.3\linewidth}
        \includegraphics[width=1\linewidth]{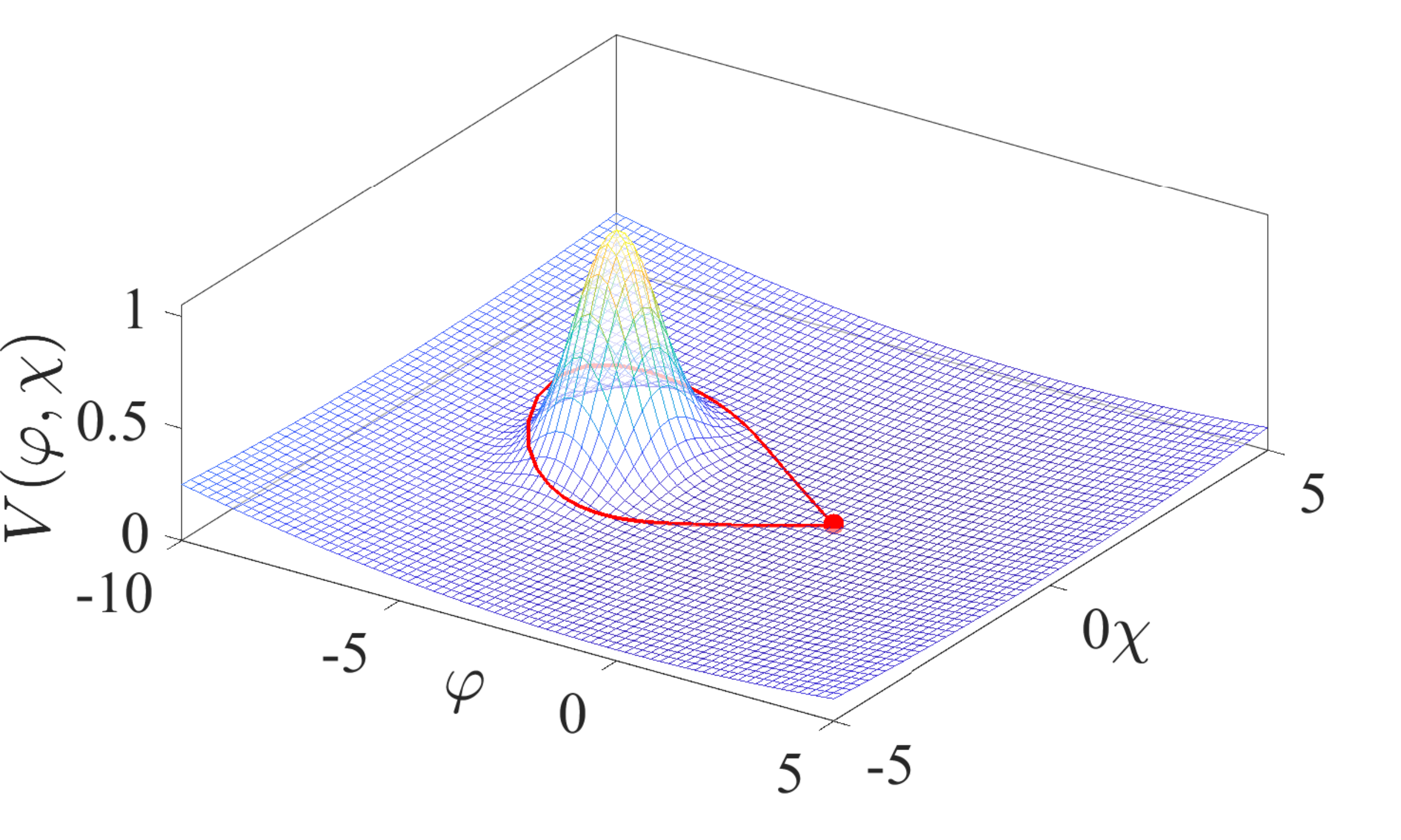}
        \caption{}
        \label{fig:V_fin_dw}
    \end{subfigure}
    \caption{
        The potential \eqref{Potential1} and the distribution (areas in red) of the $(\varphi,\chi)$-fields for (\textbf{a}) initial, (\textbf{b}) intermediate, and (\textbf{c}) final field configurations with the final state, corresponding to the domain wall.}
    \label{fig:V.solutions.dw}
\end{figure}

\begin{figure}[H]
    \begin{subfigure}[h]{0.3\linewidth}
        \includegraphics[width=1\linewidth]{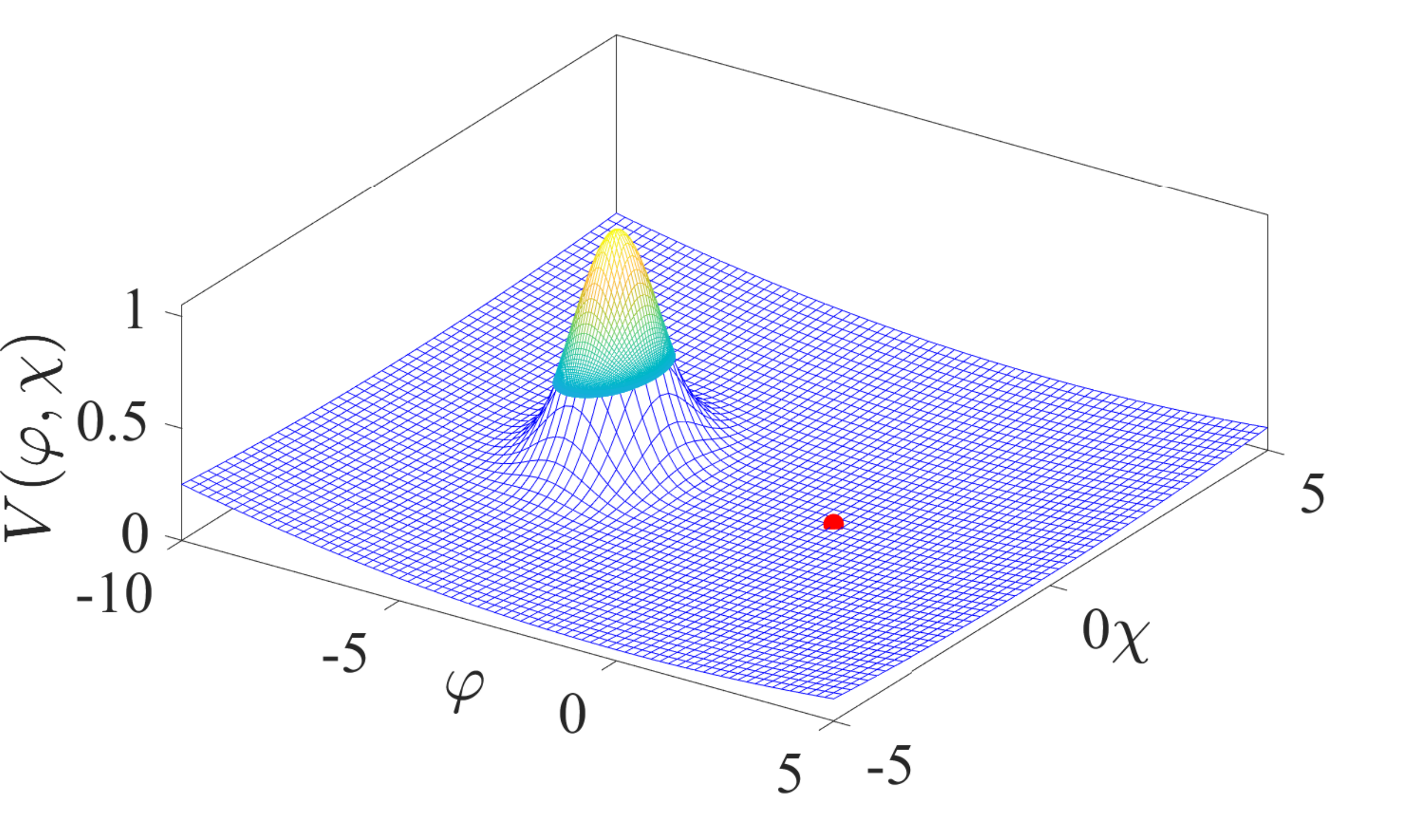}
        \caption{}
        \label{fig:V_in_str}
    \end{subfigure}
    \begin{subfigure}[h]{0.3\linewidth}
        \includegraphics[width=1\linewidth]{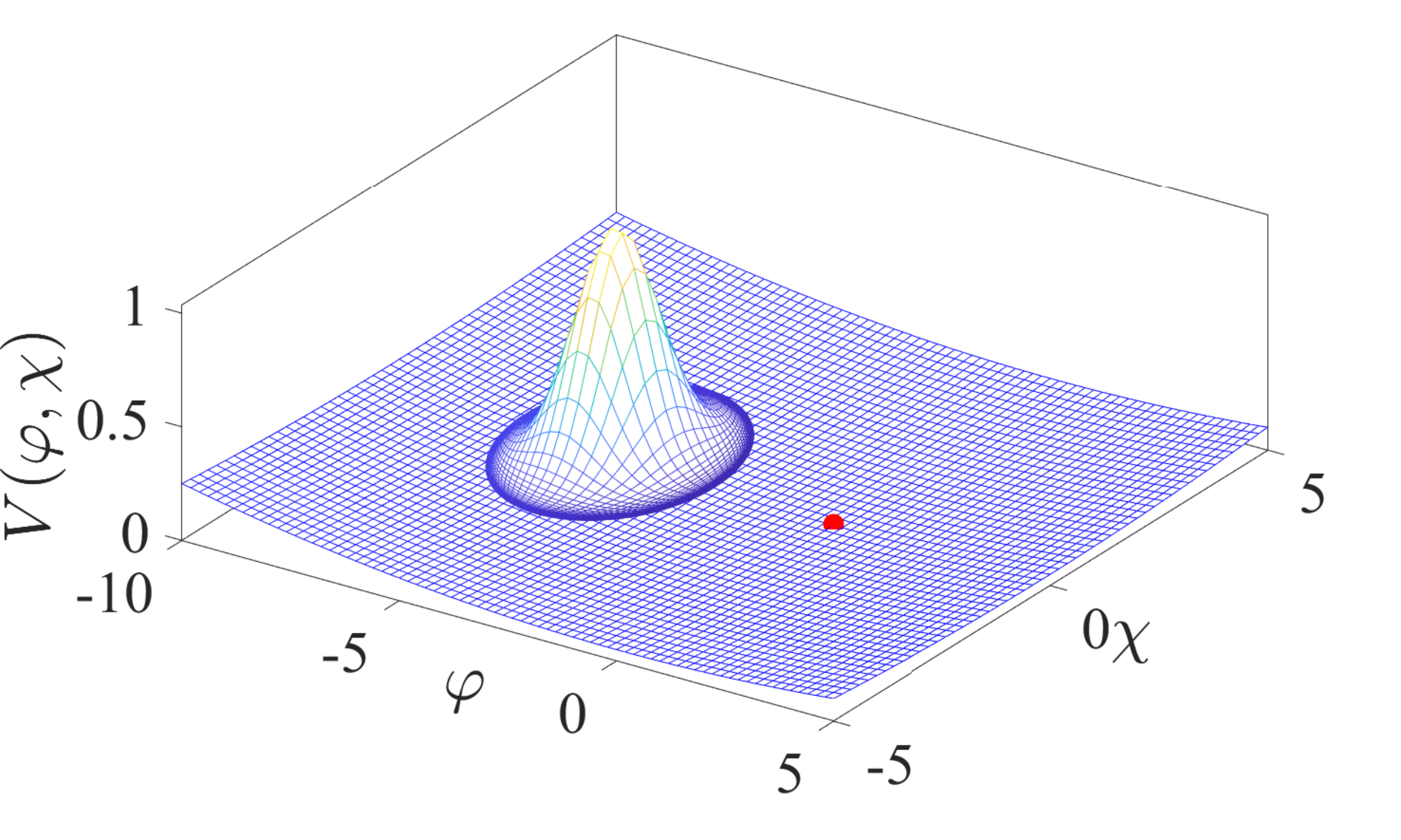}
        \caption{}
        \label{fig:V_inter_str}
    \end{subfigure}
    \begin{subfigure}[h]{0.3\linewidth}
        \includegraphics[width=1\linewidth]{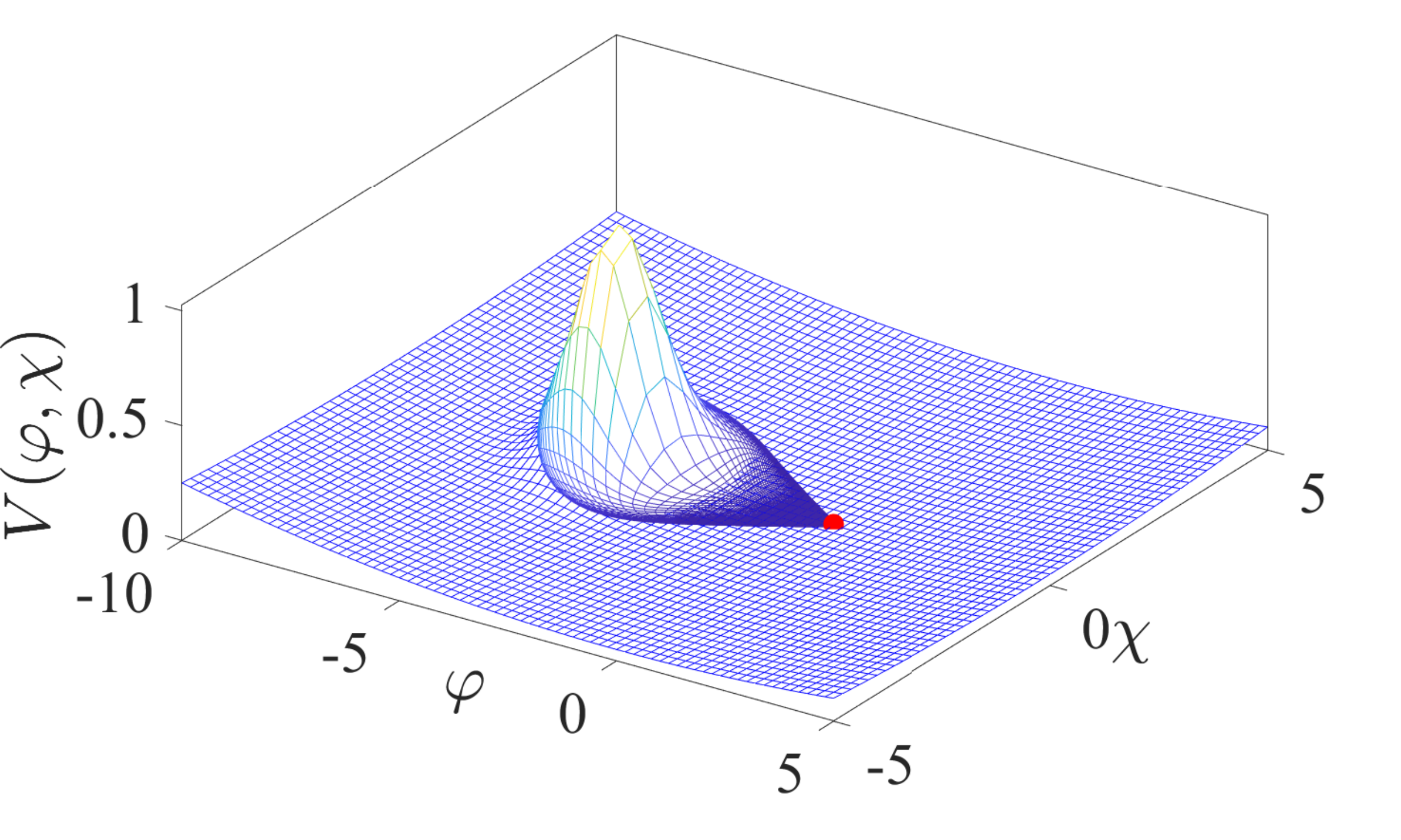}
        \caption{}
        \label{fig:V_fin_str}
    \end{subfigure}
    \caption{
        The potential \eqref{Potential1} and the distribution (areas in dark blue) of the $(\varphi,\chi)$-fields for (\textbf{a}) initial, (\textbf{b}) intermediate, and (\textbf{c}) final field configurations with  the-final state, corresponding to the string.
        }
    \label{fig:V.solutions.str}
\end{figure}

The numerical solutions of Equations \eqref{Motion} with the conditions \eqref{InitialConditions} and \eqref{BoundaryConditions} are shown in Figures~\ref{fig:V.solutions.dw} and 
\ref{fig:V.solutions.str}. For~illustration, the~following parameters have been taken: $m=0.1$, $\Lambda=\sqrt[4]{2}$, $\lambda=1$ and $(\varphi_0,\chi_0)=(-5,0)$ for the potential \eqref{Potential1} and $\mathcal{R}_0=1$, $r_0=1$ for the initial conditions \eqref{InitialConditions}.
Figures~\ref{fig:V.solutions.dw}a and \ref{fig:V.solutions.str}a show two initial field configurations at the same potential with the parameters $(\varphi_1,\chi_1) = (-8,0)$ (near the saddle point) and $(-5,0)$ (near the local maximum), respectively.  As~a result, one obtains two different field configurations corresponding to the domain wall (Figure~\ref{fig:V.solutions.dw}c) and the string-like soliton (Figure~\ref{fig:V.solutions.str}c).

\nointerlineskip

\nointerlineskip

\nointerlineskip
\begin{figure}[H]
    \begin{subfigure}[h]{0.45\linewidth}
      \includegraphics[width=1\linewidth]{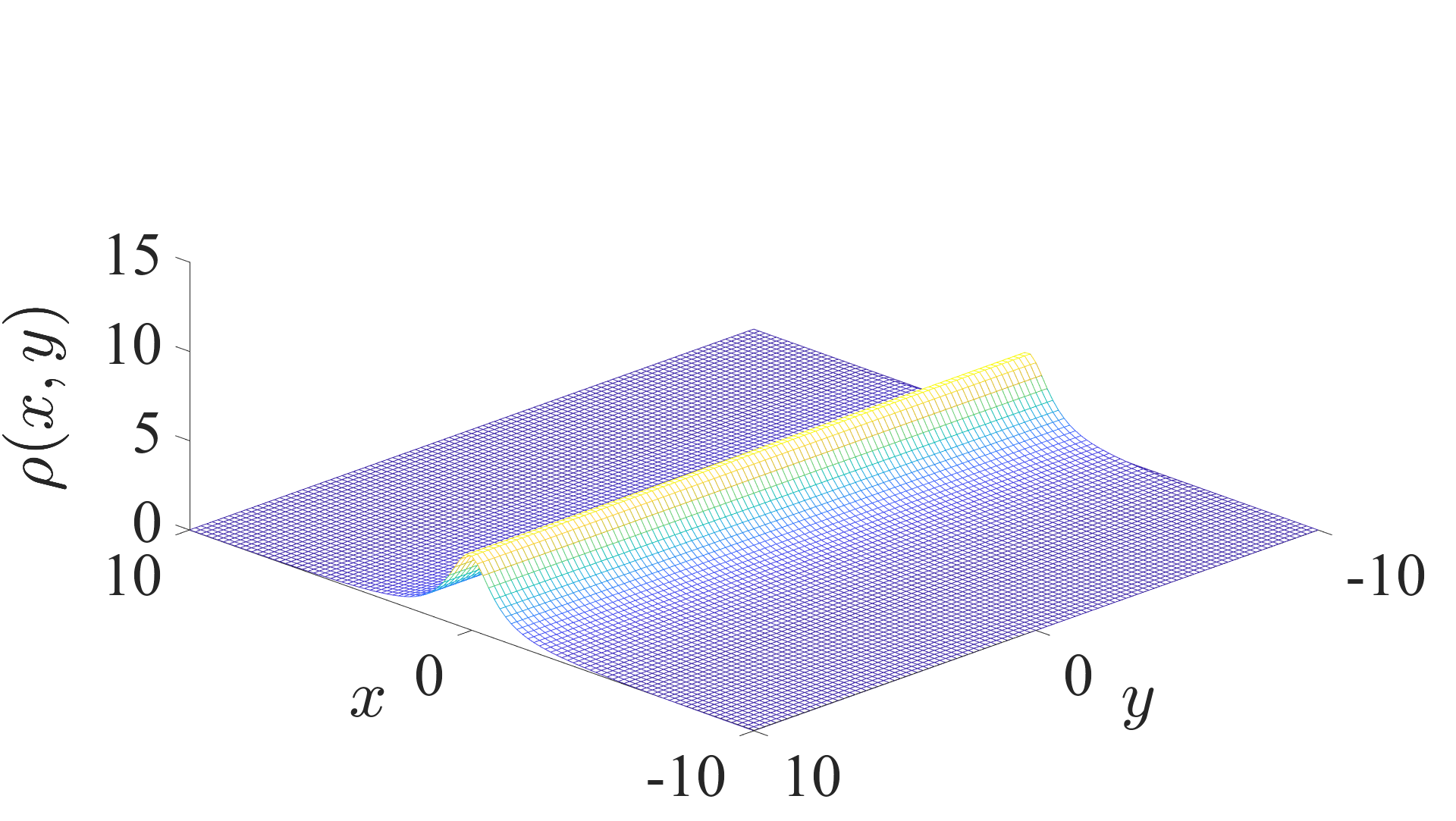} 
      \caption{}
     \label{fig:rho_fin_dw}
    \end{subfigure}
    \hfil
    \begin{subfigure}[h]{0.45\linewidth}
      \includegraphics[width=1\linewidth]{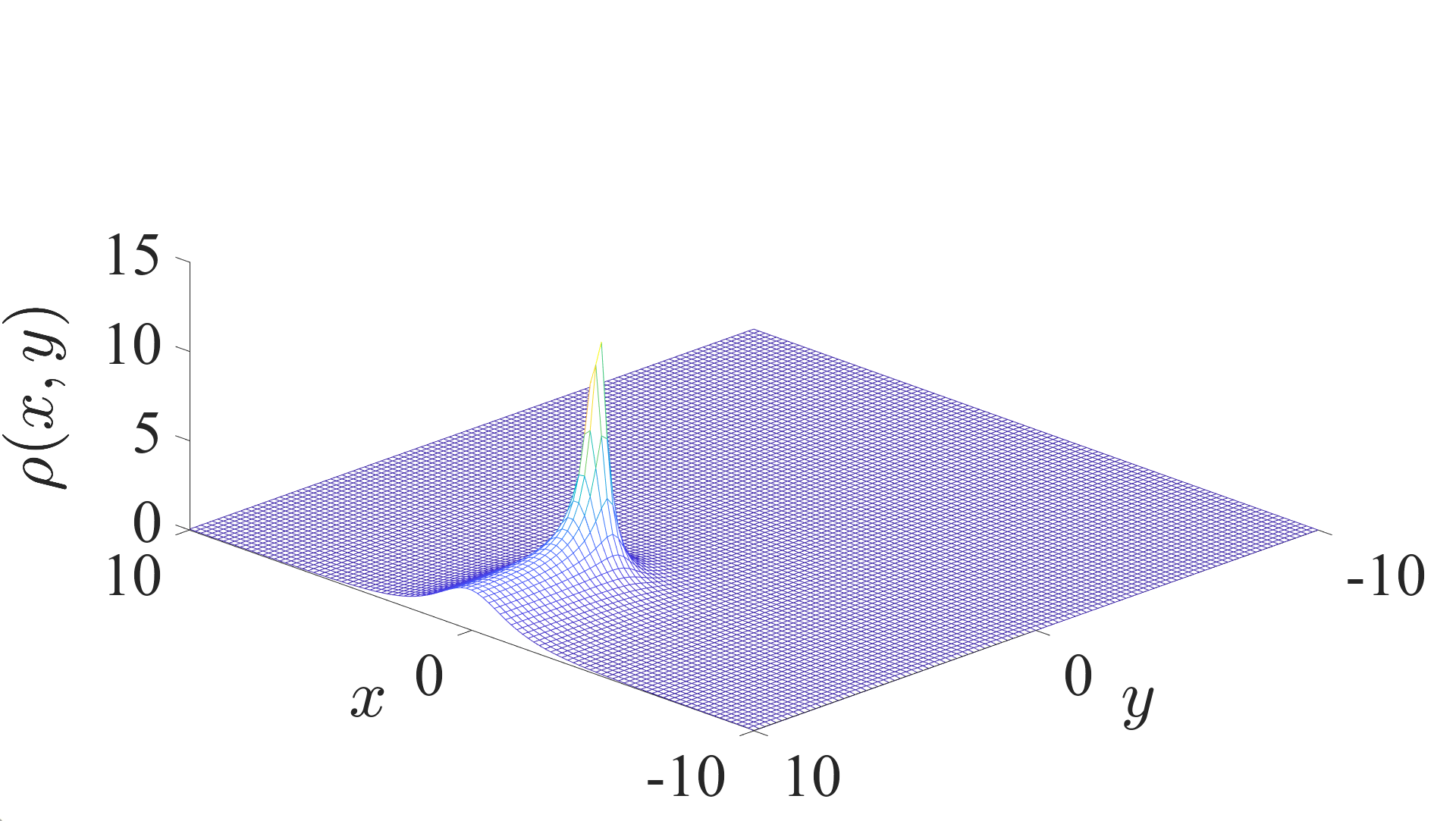}
      \caption{}
      \label{fig:rho_fin_str}
    \end{subfigure}
    \caption{
        The energy densities of the final states, corresponding to (\textbf{a}) domain wall (Figure~\ref{fig:V.solutions.dw}c), and (\textbf{b}) string with the ridge formation (Figure~\ref{fig:V.solutions.str}c).
}
    \label{energy}
\end{figure}

The energy densities of the solitons are shown in Figure~\ref{energy}. Domain walls are formed if the distribution of the initial fields is located near the saddle point of the potential, see \mbox{Figure~\ref{fig:V.solutions.dw}a.} The field values tend to the potential minimum and go around the local maximum. 
\mbox{Figure~\ref{fig:V.solutions.dw}c} demonstrates the final state; the~corresponding energy density is shown in Figure~\ref{energy}a. Likewise, string-like solutions are formed if the initial distribution covers the peak of the local maximum (Figure~\ref{fig:V.solutions.str}a). The~energy density is shown in Figure~\ref{energy}b; it differs from the known strings due to the ridge formation and motion of the formed string towards the ridge. The~ridge forms 
only if potentials have at least one saddle point and a local maximum, while vanishes for symmetric potentials. Note that  similar potential forms allow to produce the same solitons~\cite{2020arXiv201107041K}.

Both a wall and a string are formed under special (boundary) conditions. However, in~fact, the~solitons represent local manifestations (2-dimensional slices) of some entire structure in 3-dimensional space. To~study the entire structure, it is necessary to set the vacuum value of the fields as the boundary condition at infinity. Therefore, a~string ridge should smoothly pass into a domain wall. A~domain wall can be both closed and open. In the latter case, a~wall should be bounded by strings~\cite{1985PhR...121..263V}. Similarly, in~3-dimensions, the~entire structure takes the form of a bubble (for a closed wall) or a pancake (for a wall bounded by strings) \cite{1998PhRvD..59b3505C}. Thus, in~the model with the potential in question, isolated strings present only a local manifestation of an entire structure and are not formed~separately.

As it is mentioned above, the~simulation of the string formation shows that the string moves towards the ridge. Thus, the~configuration is not stable. At~the same time, this instability can be naturally explained if one considers the entire structure, a wall bounded by strings. Then, the~movement of the strings along the ridge/wall corresponds to minimizing the surface tension energy of the wall stretched between the~strings.

Let us estimate the formation probability of these structures during the 
fluctuations of the fields at the inflationary stage. As explained above, for~a careful analysis, it is necessary to consider entire 3-dimensional structures while the considered  structures are their local 2-dimensional slices. However, a~qualitative estimation helps to understand whether the structures with strings are more likely than 
ones with walls only.

\begin{figure}[H]
    \includegraphics[width=0.7\linewidth, trim={0cm 0cm 0cm 0cm}, clip]{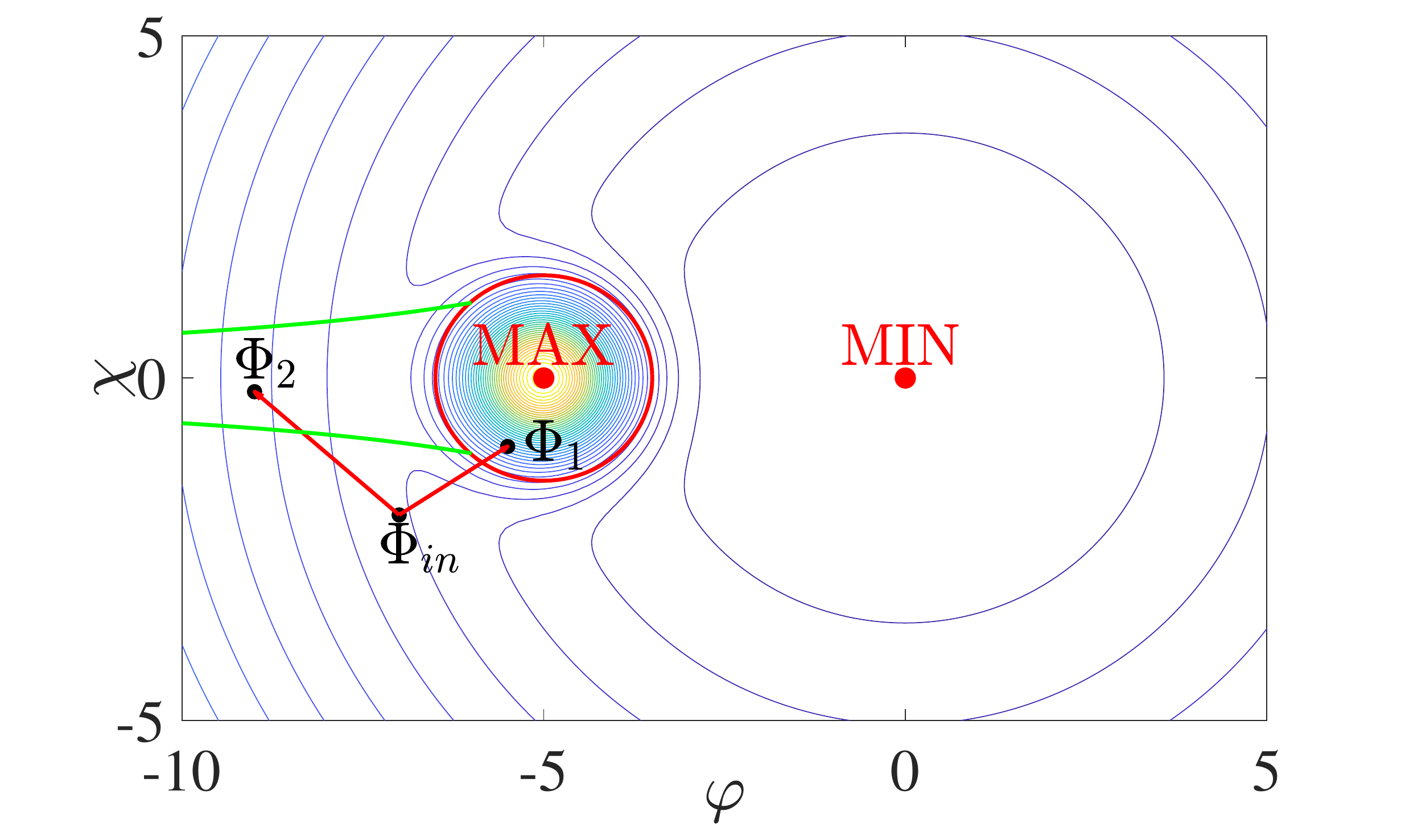}
    \caption{
        The contour plot of the potential \eqref{Potential1} of scalar fields. The areas, surrounded by the red and green curves, show the regions which might be achieved by quantum fluctuations to form strings and domain walls, respectively. The~areas differ by the factor $\varkappa\sim10^3$. If~the fluctuations lead to the field values $\Phi_{1}$ and $\Phi_{2}$, located outside the regions, the~vacuum solution inevitably occurs. The~initial possible field value, $\Phi_{\text{in}}$, and the subsequent values $\Phi_{1}$ and $\Phi_{2}$ are shown schematically. 
        The labels ``MIN'' and ``MAX'' indicate the global minimum and the local maximum (peak) of the potential, respectively.
    }
    \label{probability}
\end{figure}

To qualitatively estimate the relative probability, $P_\text{dw}$, of the production of domain walls and that of the production of strings, $P_\text{s}$, let us suppose that scalar fields have some initial values, $\Phi_{\text{in}}\equiv(\varphi_{\text{in}}, \chi_{\text{in}})$, at the beginning of the inflationary process (at an instant $t$), see Figure~\ref{probability}. At an instant $t+\delta t$, quantum fluctuations lead to the field values $\Phi_{1}$ and $\Phi_{2}$ in different causally connected spatial areas.
If the value $\Phi_1$ is close to the peak of the potential (the red area in Figure~\ref{probability}), further fluctuations are capable (with increased probability) to form the initial condition leading to the 
formation of strings like in Figure~\ref{fig:V.solutions.str}a. If the value $\Phi_2$ is close to the saddle point (the green area in Figure~\ref{probability}), the~probability of forming the initial condition necessary for the formation of walls (Figure~\ref{fig:V.solutions.dw}a) increases.

The simple geometric calculations of the ratio $\varkappa = P_\text{dw}/P_\text{s}$ to be done as follows. For~a rough estimate, one assumes that all values of the fields $\Phi(\varphi,\chi)$ in Figure~\ref{probability} are equally probable. Therefore, the~ratio between the green area in Figure~\ref{probability}, where domain walls appear, and the red area, where strings appear, shows the relative probability to form one or the other type of solutions. Both areas are numerically calculated and schematically illustrated in Figure~\ref{probability}; for~the potential parameters in question, the~ratio gives the factor $\varkappa\sim10^3$. All other positions of $\Phi_i$ lead to the vacuum solution in the entire $xy$-plane. Here, the fact that fluctuations of $\Phi$, leading to the various points of the potential $\Phi_i$, may occur with different probabilities~\cite{1982PhLB..116..335L} is neglected. Therefore, the~more accurate estimate should take into account the position of the 
starting point, $\Phi_{\text{in}}$.

Thus, the~probability $P_\text{s}$ is found to be suppressed by the factor $\varkappa\sim10^{3}$ compared to the probability $P_\text{dw}$ for the chosen potential. One has to note that $\varkappa$ does not depend on the radius $\mathcal{R}_0$ of the disk-like initial distribution and the parameter $\Lambda$, but~essentially depend on $m$. However, for~any plausible set of the parameters, $\varkappa>1$ and the production of strings is suppressed compared to the production of domain~walls.

\section{Conclusions}

In this paper, the~mechanism of soliton formation due to classical field dynamics of two real scalar fields in (2 + 1)-space-time is considered. It is shown that both domain walls and strings with ridges are possible to be formed in field models with the potential containing at least one saddle point and a local maximum; for example, the known tilted Mexican Hat potential was discussed in~\cite{2020arXiv201107041K}. Thus, even inflaton fields could lead to a formation of solitons. This to be checked in  similar models to avoid overproduction in the early Universe and possible subsequent production of primordial black holes~\cite{2000IJMPA..15.4433H, 2001JETP...92..921R, 2017JCAP...12..044D}. 

The qualitative estimates of the soliton formation probability 
are made and it is shown that the production of strings is suppressed compared to the production of domain walls. The~discussed field configurations are actually 2-dimensional spatial slices of 3-dimensional structures. Moreover, the strings with ridges and domain walls are shown to be just local manifestations of entire structures. More accurate analysis of probabilities needs to be made for entire 3-dimensional structures; this point to be considered in future.

\section*{Acknowledgements}
We are grateful to S.~G.~Rubin, K.~M.~Belotsky and V.~A.~Gani for useful discussions.
This research was funded by the Russian Foundation for Basic Research, Project №~19-02-00930.

\sloppy

\printbibliography

\end{document}